\documentstyle[amssymb,aps,epsf]{revtex}

\begin{document}
\title{Casimir energy and black hole pair creation in Schwarzschild-de Sitter
spacetime}
\author{Remo Garattini}
\address{Facolt\`{a} di Ingegneria, Universit\`{a} degli Studi di Bergamo,\\
Viale Marconi, 5, 24044 Dalmine (Bergamo) Italy.\\
E-mail: Garattini@mi.infn.it}
\maketitle

\begin{abstract}
Following the subtraction procedure for manifolds with boundaries,
we calculate by variational methods, the Schwarzschild-de Sitter
and the de Sitter space energy difference. By computing the one
loop approximation for TT tensors we discover the existence of an
unstable mode even for the non-degenerate case. This result seems
to be in agreement with the sub-maximal black hole pair creation
of Bousso-Hawking. The instability can be eliminated by the
boundary reduction method. Implications on a foam-like space are
discussed.
\end{abstract}

\section{Introduction}

An intriguing property of quantum physics is the particle creation
generated by external fields, like a constant electric or magnetic
field, or by quantum fluctuation of the vacuum. In this last case
only virtual particles are involved. However when the energy scale
is large enough virtual particles can be transformed into real. In
line of principle the same mechanism can be shared by the
gravitational field where virtual black holes \cite
{MannRoss,Hawking} can be created and annihilated in analogy with
particle physics. This particular phenomenon has been investigated
in different contexts and in particular when a cosmological
constant is introduced\cite {MannRoss,BoHaw}. In this example the
process is mediated by the corresponding gravitational instanton,
and the semiclassical nucleation rate for a pair on a given
background is given by
\begin{equation}
\Gamma =A\exp \left[ -\left( I_{inst}-I_{back}\right) \right] .
\end{equation}
$I_{inst}$ is the classical action of the gravitational instanton mediating
the pair creation, $I_{back}$ is the action of the background field, and $A$
is the prefactor containing quantum corrections. For the de\ Sitter (dS)
space the quantum creation of black holes leads to the discovery of an
unstable mode in the physical sector, when one-loop approximation is
considered\cite{GP,Young,VW}. This quantum instability is related to the $%
S^{2}\times S^{2}$ instanton responsible for the pair creation process. This
instanton, termed the Nariai instanton\cite{Nariai}, is nothing but the
extreme Schwarzschild-de Sitter (SdS) solution written in another system of
coordinates. This instability leads to spontaneous nucleation of black holes
signaling a transition from a false vacuum to a true one\cite{Coleman}. This
transition is possible when the energy stored in the boundaries is the same
for both spaces\cite{Witten}. However as remarked in Ref.\cite{VW}, the
nucleation appears with a temperature $T_{pair}=\frac{\sqrt{\Lambda _{c}}}{%
2\pi }$ different from the temperature of the heat bath, which is
the dS space with $T_{dS}=\frac{1}{2\pi }\sqrt{\frac{\Lambda
_{c}}{3}}$. This does not happen, for example, in the hot
Minkowski space where the nucleated black hole has the same
temperature as the heat bath\cite{GPY}. The same situation holds
even when we consider a negative cosmological constant, i.e.
Anti-de Sitter (AdS) space. In fact to spontaneously nucleate a
black hole, which has an intrinsic temperature $T_{S-AdS}$, the
same temperature has to be imposed to the AdS space
\cite{HawPage,Mann,Prestidge}. However in Ref.\cite {Remo}, we
have shown that a semi-classical instability (WKB) appears for
Minkowski space even at zero temperature, provided that boundary
conditions be energy preserving. The same semi-classical
instability appears also for the AdS space\cite{Remo1} with the
same energy condition. An interesting common feature between these
cases comes from the expression of the mixed Ricci tensor
$R_{i}^{a}$ computed in these different backgrounds, namely the
Schwarzschild, S-AdS and SdS metrics respectively. Indeed in the
first case, $R_{i}^{a}$ has components:
\begin{equation}
R_{i}^{a}=\left\{ -\frac{2MG}{r^{3}},\frac{MG}{r^{3}},\frac{MG}{r^{3}}%
\right\} ,
\end{equation}
while the case with the cosmological constant is
\begin{equation}
R_{i}^{a}=\left\{ -\frac{2MG}{r^{3}}\pm 2/b^{2},\frac{MG}{r^{3}}\pm 2/b^{2},%
\frac{MG}{r^{3}}\pm 2/b^{2}\right\}
\end{equation}
where the upper case and the lower case are related to the SdS and the S-AdS
metrics respectively and $b^{2}=3/\Lambda _{c}$. $\Lambda _{c}$ is the
positive cosmological constant. It is straightforward to note that the only
difference between these different tensors is in the presence of the
cosmological term. This means that the source of instability residing in the
first component appears even in the other cases. However, only at one loop
is possible to reveal the presence or the absence of such an instability. We
recall that one loop computations of the energy in this context represent a
Casimir-like energy which measures vacuum fluctuations. Following Refs.\cite
{Remo,Remo1}, we will consider a constant time slice $\Sigma $ of the SdS
manifold ${\cal M}$\footnote{%
In Appendix \ref{app1}, we will report the details concerning the
Kruskal-Szekeres description of the SdS manifold.}, whose perturbations at $%
\Sigma $ in absence of matter fields define quantum fluctuations
of the Einstein-Rosen bridge. Indeed, as will follow in Section
\ref{p1} even though the SdS is not asymptotically flat, the
hypersurface $\Sigma $ defines a wormhole with topology
$S^{2}\times I$, where $I\subset R$ is a sub-interval of $R$. This
is a consequence of having a cosmological radius which sets an
upper bound to the radial coordinate. To this purpose we will fix
our attention on a Hamiltonian with boundary
\begin{equation}
H_{T}=H_{\Sigma }+H_{\partial \Sigma }=\int_{\Sigma }d^{3}x(N{\cal H+}N_{i}%
{\cal H}^{i})+H_{\partial \Sigma },
\end{equation}
where $N$ is called the {\it lapse} function, $N_{i}$ is the {\it shift }%
function and
\begin{equation}
\left\{
\begin{array}{l}
{\cal H}{\bf =}G_{ijkl}\pi ^{ij}\pi ^{kl}\left( \frac{16\pi G}{\sqrt{g}}%
\right) -\left( \frac{\sqrt{g}}{16\pi G}\right) \left( R^{\left( 3\right) }-%
\frac{6}{b^{2}}\right) \\
{\cal H}^{i}=-2\pi _{|j}^{ij}
\end{array}
.\right.
\end{equation}
$H_{\partial \Sigma }$ represents the energy stored into the boundary. The
aim of this paper is the evaluation of
\begin{equation}
E^{SdS}\left( M,b\right) =E^{dS}\left( b\right) +\Delta E_{dS}^{SdS}\left(
M,b\right) _{|classical}+\Delta E_{dS}^{SdS}\left( M,b\right) _{|1-loop},
\label{i0}
\end{equation}
representing the total energy computed to one-loop in a SdS background. $%
E^{dS}\left( b\right) $ is the reference space energy, i.e. the de Sitter
space. $\Delta E_{dS}^{SdS}\left( M,b\right) _{|classical}$ is the energy
difference between the SdS and the dS metrics, stored in the boundaries and $%
\Delta E_{dS}^{SdS}\left( M,b\right) _{|1-loop}$ is the quantum
correction to the classical term. The rest of the paper is
structured as follows, in section \ref{p1} we compute the
quasilocal energy for the SdS space, in section \ref{p2} we give
some of the basic rules to perform the functional integration to
evaluate the energy density of the Hamiltonian approximated up to
second order in the SdS background, in section \ref{p3} we look
for stable modes of the spin-two operator acting on transverse
traceless tensors, in section \ref{p4} we show the existence of
only one negative mode under suitable conditions and we compute
the energy density for stable modes, in section \ref{N}, we
confirm the existence of one negative mode for the extreme SdS
background, namely the Nariai metric, also in our framework and we
give a computation for the stable part in analogy with its
non-extreme sector, in section \ref{p5} we find a critical radius
below which we have a stabilization of the system. We summarize
and conclude in section \ref{p6}.

\section{Quasilocal Energy for the SdS space}

\label{p1}
\begin{figure}[tbh]
\vbox{\hfil\epsfxsize=4.5cm\epsfbox{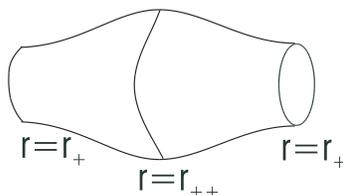}\hfil}
\caption{The geometry of the constant time slice embedded in flat
space with a polar angle suppressed. Isometric copies of this
surface can be smoothly joined at the throats, producing a
periodic $S^{2}\times R^{1}$ spatial topology.} \label{f1}
\end{figure}

In this section we fix our attention to the classical part of Eq.$%
\left( \ref{i0}\right) $. We begin to define the line element
\begin{equation}
ds^{2}=-f\left( r\right) dt^{2}+f\left( r\right) ^{-1}dr^{2}+r^{2}d\Omega
^{2},  \label{p11aa}
\end{equation}
referred to the SdS metric, where
\begin{equation}
f\left( r\right) =\left( 1-\frac{2MG}{r}-\frac{r^{2}}{b^{2}}\right) .
\label{p11a}
\end{equation}
For $\Lambda _{c}=0$ the metric describes the Schwarzschild metric, while
for $M=0$, we obtain the de Sitter metric (dS)
\begin{equation}
ds^{2}=-\left( 1-\frac{r^{2}}{b^{2}}\right) dt^{2}+\left( 1-\frac{r^{2}}{%
b^{2}}\right) ^{-1}dr^{2}+r^{2}d\Omega ^{2}.
\end{equation}
The gravitational potential $f\left( r\right) $ admits three real roots. One
is negative and it is located at
\begin{equation}
r_{-}=\frac{2}{\sqrt{3}}b\cos \left( \frac{\theta +2\pi }{3}\right) ,
\end{equation}
while
\begin{equation}
r_{+}=\frac{2}{\sqrt{3}}b\cos \left( \frac{\theta +4\pi }{3}\right) ,\
r_{++}=\frac{2}{\sqrt{3}}b\cos \left( \frac{\theta }{3}\right)
\end{equation}
are associated to the black hole and cosmological horizons respectively,
with
\begin{equation}
\cos \theta =-3MG\sqrt{3}/b.  \label{p11}
\end{equation}
However in the wormhole language, we will say that $r_{+}$ is the inner
throat and $r_{++}$ is the outer throat. Note also that the hypersurface $%
\Sigma $ is described by the three-dimensional wormhole whose metric is
\begin{equation}
ds^{2}=f\left( r\right) ^{-1}dr^{2}+r^{2}d\Omega ^{2},
\end{equation}
where $f\left( r\right) $ is given by Eq.$\left( \ref{p11a}\right) $. A
relation between the three roots is given by
\begin{equation}
\left\{
\begin{array}{c}
b^{2}=r_{+}^{2}+r_{+}r_{++}+r_{++}^{2} \\ 2Ml_{p}^{2}b^{2}=\left(
r_{+}r_{++}\right) \left( r_{+}+r_{++}\right) \\
0=r_{-}+r_{+}+r_{++}
\end{array}
.\right.  \label{p12a}
\end{equation}
Thus , we can write
\begin{equation}
f\left( r\right) =-\frac{1}{rb^{2}}\left( r-r_{+}\right) \left(
r-r_{++}\right) \left( r+r_{+}+r_{++}\right) ,  \label{p12}
\end{equation}
with
\begin{equation}
r_{+}\leq r\leq r_{++}\qquad \theta \in \left[ \frac{\pi }{2},\frac{3\pi }{2}%
\right] .  \label{p12aa}
\end{equation}
Since $r_{+}$ is a monotonic increasing function of $\theta $, while $r_{++}$
is a monotonic decreasing with
\begin{equation}
\left\{
\begin{array}{c}
r_{+}^{\upharpoonright }\in \left[ 0,b\right] \\
r_{++}^{\downharpoonright }\in \left[ 0,b\right]
\end{array}
,\right.
\end{equation}
in order to have the inequality $\left( \ref{p12aa}\right) $ preserved, we
have to consider $\theta \in \left[ \frac{\pi }{2},\pi \right] $. Indeed
when $\theta \in \left[ \pi ,\frac{3\pi }{2}\right] $ the inequality $\left(
\ref{p12aa}\right) $ is reversed and the meaning of the internal and
external roots is exchanged. Thus the de Sitter region delimited by the
bound of Eq.$\left( \ref{p12aa}\right) $, at time fixed, can be represented
as in Fig.\ref{f1}.

\begin{figure}[tbh]
\vbox{\hfil\epsfxsize=3.5cm\epsfbox{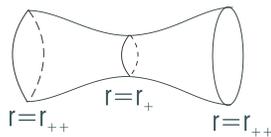}\hfil}
\caption{The same representation of Fig.\ref{f1} but with the cosmological ``%
{\it wormhole mouths}'' placed in antipodal region of the de Sitter universe
with a periodic $S^{2}\times R^{1}$ spatial topology.}
\label{f2}
\end{figure}

A common value is reached when $\theta =\pi $ where $r_{+}=r_{++}=b/\sqrt{3}$
and the metric is termed extreme. This particular case will be discussed in
section \ref{N}. However, instead of looking at the de Sitter region with
the topology of Fig.\ref{f1}, we will look at the Einstein-Rosen bridge
corresponding to the inner bifurcation surface depicted in Fig.\ref{f2}.

At this point we can discuss the computation of the classical energy term
\begin{equation}
E^{SdS}\left( M,b\right) =E^{dS}\left( b\right) +\Delta E_{dS}^{SdS}\left(
M,b\right) _{|classical},
\end{equation}
which can be computed by means of quasilocal energy. Quasilocal energy is
defined as the value of the Hamiltonian that generates unit time
translations orthogonal to the two-dimensional boundary,
\begin{equation}
\Delta E_{dS}^{SdS}\left( M,b\right) _{|classical}=\frac{1}{8\pi G}%
\int_{S^{2}}d^{2}x\sqrt{\sigma }\left( k-k^{0}\right) ,
\end{equation}
where $\left| N\right| =1$ at $S^{2}$ and $k^{0}$ is the trace of the
extrinsic curvature corresponding to the reference space, which in this case
is the dS space. For practical purposes, however, it is convenient to embed
both spaces (Sds and dS) into flat space and perform the subtraction
procedure. To this purpose the radial coordinate $x$ continuous on ${\cal M}$
is defined by
\begin{equation}
dx=\pm \frac{dr}{\sqrt{1-\frac{2MG}{r}-\frac{r^{2}}{b^{2}}}},  \label{p13}
\end{equation}
where the plus sign is relative to $\Sigma _{+}$, while the minus
sign is related to $\Sigma _{-}$. The surfaces located at $r_{+}$
and $r_{++}$ are bifurcation surfaces denoted $S_{+}^{0}$ and
$S_{++}^{0}$, respectively. When $M=0$, we obtain the embedding of
dS space into flat space. In $\Sigma _{+}$ the evaluation of
$\Delta E_{dS}^{SdS}\left( M,b\right) _{|classical}$ can be
obtained as follows: first we consider the static Einstein-Rosen
bridge associated to the SdS space\cite{FroMar,HawHor}
\begin{equation}
ds^{2}=-N^{2}\left( r\right) dt^{2}+g_{xx}dx^{2}+r^{2}\left( x\right)
d\Omega ^{2},  \label{p14}
\end{equation}
where $N$, $g_{xx}$, and $r$ are functions of $x$ defined by Eq.$\left( \ref
{p13}\right) $. Second, we consider the boundary $S_{+}^{2}$, located at $%
x\left( r\right) =\bar{x}^{+}\left( R\right) $, and its associated normal $%
n^{\mu }=\left( h^{xx}\right) ^{\frac{1}{2}}\delta _{y}^{\mu }$. The
expression of the trace
\begin{equation}
k=-\frac{1}{\sqrt{h}}\left( \sqrt{h}n^{\mu }\right) _{,\mu },
\end{equation}
gives for the SdS space
\begin{equation}
k^{SdS}=-2\frac{r,_{x}}{r}_{|SdS}=-2\frac{\sqrt{f\left( r\right) }}{r}%
_{|SdS}=-\frac{2}{r}\sqrt{1-\frac{2MG}{r}-\frac{r^{2}}{b^{2}}}.  \label{p15}
\end{equation}
Note that if we make the identification $N^{2}=1-\frac{2MG}{r}-\frac{r^{2}}{%
b^{2}}$, the line element $\left( \ref{p14}\right) $ reduces to the SdS
metric written in another form. The same applies to the dS metric by putting
$M=0$. Nevertheless for our purposes the form of $N\left( r\right) $ can be
left unspecified. Thus the computation of $E_{+}$ gives
\begin{equation}
\Delta E_{dS}^{SdS}\left( M,b\right) _{|classical}=\frac{1}{8\pi G}%
\int_{S^{2}}d\Omega ^{2}r^{2}\left[ \frac{-2\sqrt{f\left( r\right) }}{r}+%
\frac{2\sqrt{f\left( r\right) _{|M=0}}}{r}\right] _{|r=R}
\end{equation}
\begin{equation}
=-\frac{R}{G}\left[ \sqrt{1-\frac{2MG}{R}-\frac{R^{2}}{b^{2}}}-\sqrt{1-\frac{%
R^{2}}{b^{2}}}\right] ,
\end{equation}
where we have set $M=0$ in $k^{SdS}$ to obtain the dS energy contribution.
When $R\gg b$, $\Delta E_{dS}^{SdS}\left( M,b\right) _{|classical}\simeq
-iMb/R$. Thus for every finite value of the boundary exceeding the
cosmological radius, the classical energy acquires an imaginary component
which will not be here considered\footnote{%
To deal with this case it is better to introduce the quasilocal mass,
defined as
\[
\Delta M_{dS}^{SdS}\left( M,b\right) _{|classical}=\frac{1}{8\pi G}%
\int_{S^{2}}d^{2}xN\sqrt{\sigma }\left( k-k^{0}\right)
\]
\[
\mathrel{\mathop{\simeq }\limits_{R\gg b}}%
\left( -iM\frac{b}{R}\right) \left( i\frac{R}{b}\right) =M.
\]
}. In contrast, if we consider the approximation $R/b\ll 1$ and $%
2MG/R\ll 1$\cite{Maluf}, we obtain
\begin{equation}
\Delta E_{dS}^{SdS}\left( M,b\right) _{|classical}\simeq -\frac{R}{G}\left[
\left( 1-\frac{MG}{R}-\frac{R^{2}}{2b^{2}}\right) -\left( 1-\frac{R^{2}}{%
2b^{2}}\right) \right] =M.
\end{equation}
Thus the energy contribution limited to $\Sigma _{+}$ gives
\begin{equation}
E^{SdS}\left( M,b\right) =E^{dS}\left( b\right) +\Delta E_{dS}^{SdS}\left(
M,b\right) _{|classical}^{+}=E^{dS}\left( b\right) +M,
\end{equation}
that is the dS space cannot decay into the SdS space because the associated
boundary energy is different. This is in complete analogy with the
Schwarzschild and the S-AdS cases. However if we look at the whole
hypersurface $\Sigma =\Sigma _{+}\cup \Sigma _{-}$ the total classical
energy becomes
\[
E^{SdS}\left( M,b\right) =E^{dS}\left( b\right) +E_{tot}\left( M,b\right)
\]
\begin{equation}
=E^{dS}\left( b\right) +\Delta E_{dS}^{SdS}\left( M,b\right)
_{|classical}^{+}+\Delta E_{dS}^{SdS}\left( M,b\right) _{|classical}^{-}
\end{equation}
with
\[
\Delta E_{dS}^{SdS}\left( M,b\right) _{|classical}^{+}=\frac{1}{8\pi G}%
\int_{S_{+}^{2}}d^{2}x\sqrt{\sigma }\left( k-k^{0}\right) ,
\]
\begin{equation}
\Delta E_{dS}^{SdS}\left( M,b\right) _{|classical}^{-}=-\frac{1}{8\pi G}%
\int_{S_{-}^{2}}d^{2}x\sqrt{\sigma }\left( k-k^{0}\right) .
\end{equation}
Here the boundaries $S_{+}^{2}$ and $S_{-}^{2}$ are located in the two
disconnected regions ${\cal M}_{+}$ and ${\cal M}_{-}$ respectively with
coordinate values $x=\bar{x}^{\pm }$ and the trace of the extrinsic
curvature in both regions is
\begin{equation}
k^{SdS}=\left\{
\begin{array}{c}
-2r,_{x}/r\qquad on\ \Sigma _{+} \\
2r,_{x}/r\qquad on\ \Sigma _{-}
\end{array}
\right. .
\end{equation}
Thus one gets
\begin{equation}
\Delta E_{dS}^{SdS}\left( M,b\right) _{|classical}^{\pm }=\left\{
\begin{array}{c}
M\qquad on\ S_{+}^{2} \\
-M\qquad on\ S_{-}^{2}
\end{array}
\right. ,
\end{equation}
where for $E_{-}$ we have used the conventions relative to $\Sigma _{-}$ and
$S_{-}^{2}$. Therefore for every value of the boundary $R$, (provided we
take symmetric boundary conditions with respect to the bifurcation surface),
we have
\begin{equation}
E^{SdS}\left( M,b\right) =E^{dS}\left( b\right) +M+\left( -M\right)
=E^{dS}\left( b\right) ,
\end{equation}
namely the energy is conserved. As stressed in Ref.\cite{FroMar}, since we
have a spacetime with a bifurcation surface, the quantities $\Delta
E_{dS}^{SdS}\left( M,b\right) _{|classical}^{+}$ and $\Delta
E_{dS}^{SdS}\left( M,b\right) _{|classical}^{-}$ have the same relative
sign, while the total energy is given by the sum $\Delta E_{dS}^{SdS}\left(
M,b\right) _{|classical}^{+}+\Delta E_{dS}^{SdS}\left( M,b\right)
_{|classical}^{-}$\footnote{%
In Ref.\cite{FroMar} we have a subtraction instead of a sum. This is due to
conventions adopted.}. The energies associated to the boundaries are
symmetric and they have the same relative sign while the total energy
reflects the orientation reversal of the two boundaries. Since the total
classical energy is conserved we can discuss the existence of an
instability. To this aim we refer to the variational approach to compute the
energy density to one-loop\cite{Remo,Remo1,CJT,CP,Remo2}.

\section{Energy Density Calculation in Schr\"{o}dinger Representation}

\label{p2}In previous section we have fixed our attention to the classical
part of Eq.$\left( \ref{i0}\right) $. In this section, we apply the same
calculation scheme of Refs.\cite{Remo,Remo1} to compute one loop corrections
to the classical SdS term. Like the Schwarzschild and the S-AdS case, there
appear two classical constraints
\begin{equation}
\left\{
\begin{array}{l}
{\cal H}\text{ }=0 \\
{\cal H}^{i}=0
\end{array}
\right.
\end{equation}
and two {\it quantum} constraints
\begin{equation}
\left\{
\begin{array}{l}
{\cal H}\tilde{\Psi}\text{ }=0 \\
{\cal H}^{i}\tilde{\Psi}=0
\end{array}
\right.
\end{equation}
for the Hamiltonian respectively, which are satisfied both by the SdS and dS
metric on shell. ${\cal H}\tilde{\Psi}$ $=0$ is known as the {\it %
Wheeler-DeWitt} equation (WDW). Our purpose is the computation of
\begin{equation}
\Delta E_{dS}^{SdS}\left( M,b\right) _{|1-loop}=\frac{\left\langle \Psi
\left| H_{\Sigma }^{SdS}-H_{\Sigma }^{dS}\right| \Psi \right\rangle }{%
\left\langle \Psi |\Psi \right\rangle }  \label{p21}
\end{equation}
where $H_{\Sigma }^{SdS}$ and $H_{\Sigma }^{dS}$ are the total Hamiltonians
referred to the SdS and dS spacetimes respectively for the volume term\cite
{Remo} and $\Psi $ is a wave functional obtained following the usual WKB
expansion of the WDW solution\footnote{%
See also\cite{SusUgl,DLM,Romeo} for other applications of the WKB
approximation concerning black hole physics.}. In this context,
the approximated wave functional will be substituted by a {\it
trial wave functional} of the gaussian form according to the
variational approach we shall use to evaluate $\Delta
E_{dS}^{SdS}\left( M,b\right) _{|1-loop}$. To compute such a
quantity we will consider on $\Sigma $ deviations from the SdS
metric spatial section of the form,
\begin{equation}
g_{ij}=\bar{g}_{ij}+h_{ij},
\end{equation}
where
\begin{equation}
\bar{g}_{ij}dx^{i}dx^{j}=\left( 1-\frac{2MG}{r}-\frac{r^{2}}{b^{2}}\right)
^{-1}dr^{2}+r^{2}d\Omega ^{2}
\end{equation}
is the spatial SdS background. By setting $M=0$ in Eq.$\left( \ref{p11a}%
\right) $ on the same slice we will obtain perturbations also for the de
Sitter metric. Thus the expansion of the three-scalar curvature $\int d^{3}x%
\sqrt{g}R^{\left( 3\right) }$ up to $o\left( h^{2}\right) $ gives
\[
\int_{\Sigma }d^{3}x\sqrt{\bar{g}}\left[ -\frac{1}{4}h\triangle h+\frac{1}{4}%
h^{li}\triangle h_{li}-\frac{1}{2}h^{ij}\nabla _{l}\nabla _{i}h_{j}^{l}+%
\frac{1}{2}h\nabla _{l}\nabla _{i}h^{li}-\frac{1}{2}h^{ij}R_{ia}h_{j}^{a}+%
\frac{1}{2}hR_{ij}h^{ij}\right]
\]
\begin{equation}
+\int_{\Sigma }d^{3}x\sqrt{\bar{g}}\left[ \frac{1}{4}h^{li}\left( R^{\left(
0\right) }-6/b^{2}\right) h_{li}-\frac{1}{4}h^{li}\left( R^{\left( 0\right)
}\right) h_{li}+\frac{1}{4}h\left( R^{\left( 0\right) }\right) h\right] ,
\end{equation}
where $R^{\left( 0\right) }$ is the three-scalar curvature on-shell. To
explicitly make calculations, we need an orthogonal decomposition for both $%
\pi _{ij\text{ }}$and $h_{ij}$ to disentangle gauge modes from physical
deformations. We define the inner product

\begin{equation}
\left\langle h,k\right\rangle :=\int_{\Sigma }\sqrt{g}G^{ijkl}h_{ij}\left(
x\right) k_{kl}\left( x\right) d^{3}x,
\end{equation}
by means of the inverse WDW metric $G_{ijkl}$, to have a metric on the space
of deformations, i.e. a quadratic form on the tangent space at h, with
\begin{equation}
G^{ijkl}=(g^{ik}g^{jl}+g^{il}g^{jk}-2g^{ij}g^{kl}).
\end{equation}
The inverse metric is defined on co-tangent space and it assumes the form
\begin{equation}
\left\langle p,q\right\rangle :=\int_{\Sigma }\sqrt{g}G_{ijkl}p^{ij}\left(
x\right) q^{kl}\left( x\right) d^{3}x\text{,}
\end{equation}
so that
\begin{equation}
G^{ijnm}G_{nmkl}=\frac{1}{2}\left( \delta _{k}^{i}\delta _{l}^{j}+\delta
_{l}^{i}\delta _{k}^{j}\right) .
\end{equation}
Note that in this scheme the ``inverse metric'' is actually the WDW metric
defined on phase space. Now, we have the desired decomposition on the
tangent space of 3-metric deformations\cite{BerEbi,York}:
\begin{equation}
h_{ij}=\frac{1}{3}hg_{ij}+\left( L\xi \right) _{ij}+h_{ij}^{\bot }
\label{b0}
\end{equation}
where the operator $L$ maps $\xi _{i}$ into symmetric tracefree tensors
\begin{equation}
\left( L\xi \right) _{ij}=\nabla _{i}\xi _{j}+\nabla _{j}\xi _{i}-\frac{2}{3}%
g_{ij}\left( \nabla \cdot \xi \right) .
\end{equation}
Then the inner product between three-geometries becomes
\[
\left\langle h,h\right\rangle :=\int_{\Sigma }\sqrt{g}G^{ijkl}h_{ij}\left(
x\right) h_{kl}\left( x\right) d^{3}x=
\]
\begin{equation}
\int_{\Sigma }\sqrt{g}\left[ -\frac{2}{3}h^{2}+\left( L\xi \right)
^{ij}\left( L\xi \right) _{ij}+h^{ij\bot }h_{ij}^{\bot }\right] .  \label{b1}
\end{equation}
With the orthogonal decomposition in hand we can define the trial wave
functional
\begin{equation}
\Psi \left[ h_{ij}\left( \overrightarrow{x}\right) \right] ={\cal N}\exp
\left\{ -\frac{1}{4l_{p}^{2}}\left[ \left\langle hK^{-1}h\right\rangle
_{x,y}^{\bot }+\left\langle \left( L\xi \right) K^{-1}\left( L\xi \right)
\right\rangle _{x,y}^{\Vert }+\left\langle hK^{-1}h\right\rangle
_{x,y}^{Trace}\right] \right\} ,
\end{equation}
where ${\cal N}$ is a normalization factor. Since we are only interested in
the perturbations of the physical degrees of freedom, we will only fix our
attention on the TT (traceless and transverseless) tensor sector, therefore
reducing the previous form into
\begin{equation}
\Psi \left[ h_{ij}\left( \overrightarrow{x}\right) \right] ={\cal N}\exp
\left\{ -\frac{1}{4l_{p}^{2}}\left\langle hK^{-1}h\right\rangle _{x,y}^{\bot
}\right\} .
\end{equation}
This restriction is motivated by the fact that if an instability
appears this will be in the physical sector referred to TT
tensors, namely a nonconformal instability. This choice seems to
be corroborated by the action decomposition of\cite{GriKos}, where
only TT tensors contribute to the partition function\footnote{%
See also\cite{EKP} for another point of view.}. To calculate the
energy density, we need to know the action of some basic operators
on $\Psi \left[ h_{ij}\right] $\cite{CJT}. The action of the
operator $h_{ij}$ on $|\Psi \rangle =\Psi \left[ h_{ij}\right] $
is realized by
\begin{equation}
h_{ij}\left( x\right) |\Psi \rangle =h_{ij}\left( \overrightarrow{x}\right)
\Psi \left[ h_{ij}\right] ,
\end{equation}
while the action of the operator $\pi _{ij}$ on $|\Psi \rangle $, in
general, is
\begin{equation}
\pi _{ij}\left( x\right) |\Psi \rangle =-i\frac{\delta }{\delta h_{ij}\left(
\overrightarrow{x}\right) }\Psi \left[ h_{ij}\right] .
\end{equation}
The inner product is defined by the functional integration
\begin{equation}
\left\langle \Psi _{1}\mid \Psi _{2}\right\rangle =\int \left[ {\cal D}h_{ij}%
\right] \Psi _{1}^{\ast }\left\{ h_{ij}\right\} \Psi _{2}\left\{
h_{kl}\right\}
\end{equation}
and by applying previous functional integration rules, we obtain the
expression of the one-loop-like Hamiltonian form for TT (traceless and
transverseless) deformations\cite{Remo,Remo1}
\begin{equation}
H^{\bot }=\frac{1}{4l_{p}^{2}}\int_{{\cal M}}d^{3}x\sqrt{g}G^{ijkl}\left[
K^{-1\bot }\left( x,x\right) _{ijkl}+\left( \triangle _{2}\right)
_{j}^{a}K^{\bot }\left( x,x\right) _{iakl}\right] .  \label{p22}
\end{equation}
The propagator $K^{\bot }\left( x,x\right) _{iakl}$ comes from a functional
integration and it can be represented as
\begin{equation}
K^{\bot }\left( \overrightarrow{x},\overrightarrow{y}\right)
_{iakl}:=\sum_{N}\frac{h_{ia}^{\bot }\left( \overrightarrow{x}\right)
h_{kl}^{\bot }\left( \overrightarrow{y}\right) }{2\lambda _{N}\left(
p\right) },
\end{equation}
where $h_{ia}^{\bot }\left( \overrightarrow{x}\right) $ are the
eigenfunctions of $\triangle _{2j}^{a}$ and $\lambda _{N}\left( p\right) $
are infinite variational parameters.

\section{The Schwarzschild-de Sitter Metric spin 2 operator and the
evaluation of the energy density}

\label{p3}The Spin-two operator for the SdS metric is defined by
\begin{equation}
\left( \triangle _{2}\right) _{j}^{a}:=-\triangle \delta
_{j}^{a}+2R_{j}^{a}-6/b^{2}\delta _{j}^{a}  \label{p3a}
\end{equation}
where $\triangle $ is the curved Laplacian (Laplace-Beltrami operator) on a
SdS background and $R_{j\text{ }}^{a}$ is the mixed Ricci tensor whose
components are:
\begin{equation}
R_{i}^{a}=\left\{ -\frac{2MG}{r^{3}}+2/b^{2},\frac{MG}{r^{3}}+2/b^{2},\frac{%
MG}{r^{3}}+2/b^{2}\right\} .
\end{equation}
As stressed in the introduction, the form of the mixed Ricci tensor for the
SdS space has the same dependence for the radial coordinate for both the
Schwarzschild and the S-AdS spaces. Thus to evaluate the energy density, we
are led to study the following eigenvalue equation
\begin{equation}
\left( -\triangle \delta _{j}^{a}+2R_{j}^{a}-6/b^{2}\delta _{j}^{a}\right)
h_{a}^{i}=E^{2}h_{j}^{i}  \label{p31}
\end{equation}
where $E^{2}$ is the eigenvalue of the corresponding equation. In doing so,
we follow Regge and Wheeler in analyzing the equation as modes of definite
frequency, angular momentum and parity\cite{RW}. The quantum number
corresponding to the projection of the angular momentum on the z-axis will
be set to zero. This choice will not alter the contribution to the total
energy since we are dealing with a spherical symmetric problem. In this
case, Regge-Wheeler decomposition shows that the even-parity
three-dimensional perturbation is
\begin{equation}
h_{ij}^{even}\left( r,\vartheta ,\phi \right) =diag\left[ H\left( r\right)
\left( 1-\frac{2MG}{r}-\frac{r^{2}}{b^{2}}\right) ^{-1},r^{2}K\left(
r\right) ,r^{2}\sin ^{2}\vartheta K\left( r\right) \right] Y_{l0}\left(
\vartheta ,\phi \right) .  \label{p32}
\end{equation}
In this representation $H\left( r\right) $ and $K\left( r\right) $ behave as
they were scalar fields. For a generic value of the angular momentum $L$,
one gets
\begin{equation}
\left\{
\begin{array}{c}
\left( -\triangle _{l}-\frac{4MG}{r^{3}}-\frac{2}{b^{2}}\right) H\left(
r\right) =E_{l,H}^{2}H\left( r\right) \\
\\
\left( -\triangle _{l}+\frac{2MG}{r^{3}}-\frac{2}{b^{2}}\right) K\left(
r\right) =E_{l,K}^{2}K\left( r\right)
\end{array}
,\right.  \label{p33}
\end{equation}
where $E_{l,H}^{2}$ and $E_{l,K}^{2}$ are the eigenvalues for the $H\left(
r\right) $ field and the $K\left( r\right) $ field respectively. The
Laplacian restricted to $\Sigma $ is
\begin{equation}
\triangle _{l}=\left( 1-\frac{2MG}{r}-\frac{r^{2}}{b^{2}}\right) \frac{d^{2}%
}{dr^{2}}+\left( \frac{2r-3MG}{r^{2}}-3\frac{r}{b^{2}}\right) \frac{d}{dr}-%
\frac{l\left( l+1\right) }{r^{2}}.  \label{p33a}
\end{equation}
Defining reduced fields
\begin{equation}
H\left( r\right) =\frac{h\left( r\right) }{r};\qquad K\left( r\right) =\frac{%
k\left( r\right) }{r},
\end{equation}
and passing to the proper geodesic distance from the {\it throat} of the
bridge defined by Eq.$\left( \ref{p13}\right) $, the system $\left( \ref{p33}%
\right) $ becomes\footnote{%
The system is invariant in form if we make the minus choice in
Eq.$\left( \ref{p13}\right) $.}
\begin{equation}
\left\{
\begin{array}{c}
-\frac{d^{2}}{dx^{2}}h\left( x\right) +\left( V_{l}^{-}\left( x\right) -%
\frac{3}{b^{2}}\right) h\left( x\right) =E_{l}^{2}h\left( x\right) \\
\\
-\frac{d^{2}}{dx^{2}}k\left( x\right) +\left( V_{l}^{+}\left( x\right) -%
\frac{3}{b^{2}}\right) k\left( x\right) =E_{l}^{2}k\left( x\right)
\end{array}
\right.  \label{p34}
\end{equation}
with
\begin{equation}
V_{l}^{\mp }\left( x\right) =\frac{l\left( l+1\right) }{r^{2}\left( x\right)
}\mp \frac{3MG}{r\left( x\right) ^{3}}.
\end{equation}
When $r\longrightarrow r_{0}>r_{+}$%
\begin{equation}
x\left( r\right) \simeq \sqrt{2\kappa _{+}\left( r-r_{+}\right) }\qquad
V_{l}^{\mp }\left( x\right) \longrightarrow \frac{l\left( l+1\right) }{%
r_{0}^{2}}\mp \frac{3MG}{r_{0}^{3}}=const,
\end{equation}
where
\begin{equation}
\kappa _{+}=\lim\limits_{r\rightarrow r_{+}}\frac{1}{2}\left| g_{00}^{\prime
}\left( r\right) \right| =\frac{\left( r_{+}-r_{-}\right) \left(
r_{++}-r_{+}\right) }{2b^{2}r_{+}}  \label{p34b}
\end{equation}
is the ``inner'' surface gravity associated with the smallest root. The
solution of $\left( \ref{p34}\right) $ when $r\longrightarrow r_{0}>r_{+}$
for both backgrounds is
\begin{equation}
h\left( px\right) =k\left( px\right) =\sqrt{\frac{2}{\pi }}\sin \left(
px\right) .
\end{equation}
This choice is dictated by the requirement that
\begin{equation}
h\left( x\right) ,k\left( x\right) \rightarrow 0\text{\qquad when\qquad }%
x\left( r\right) \rightarrow x\left( r_{+}\right) \simeq 0.
\end{equation}
Thus the propagator becomes
\begin{equation}
K_{\pm }^{\bot }\left( x,y\right) =\frac{V}{2\pi ^{2}}\int_{0}^{\infty
}dpp^{2}\frac{\sin \left( px\right) }{r\left( x\right) }\frac{\sin \left(
py\right) }{r\left( y\right) }\frac{Y_{l0}\left( \vartheta ,\phi \right)
Y_{l^{\prime }0}\left( \vartheta ,\phi \right) }{\lambda _{\pm }\left(
p\right) }  \label{p35}
\end{equation}
$\lambda _{\pm }\left( p\right) $ is referred to the potential function $%
V_{l}^{\pm }\left( x\right) $. Substituting Eq.$\left( \ref{p35}\right) $ in
Eq.$\left( \ref{p22}\right) $ one gets (after normalization in spin space
and after a rescaling of the fields in such a way as to absorb $l_{p}^{2}$)
\begin{equation}
E\left( M,b,\lambda \right) =\frac{V}{8\pi ^{2}}\sum_{l=0}^{\infty
}\sum_{i=1}^{2}\int_{0}^{\infty }dpp^{2}\left[ \lambda _{i}\left( p\right) +%
\frac{E_{i}^{2}\left( p,M,b,l\right) }{\lambda _{i}\left( p\right) }\right]
\label{p36}
\end{equation}
where
\begin{equation}
E_{1,2}^{2}\left( p,M,b,l\right) =p^{2}+\frac{l\left( l+1\right) }{r_{0}^{2}}%
\mp \frac{3MG}{r_{0}^{3}}-\frac{3}{b^{2}},
\end{equation}
$\lambda _{i}\left( p\right) $ are variational parameters corresponding to
the eigenvalues for a (graviton) spin-two particle in an external field and $%
V$ is the volume of the system. By minimizing $\left( \ref{p36}\right) $
with respect to $\lambda _{i}\left( p\right) $ one obtains $\overline{%
\lambda }_{i}\left( p\right) =\left[ E_{i}^{2}\left( p,M,b,l\right) \right]
^{\frac{1}{2}}$ and
\begin{equation}
E\left( M,b,\overline{\lambda }\right) =\frac{V}{8\pi ^{2}}%
\sum_{l=0}^{\infty }\sum_{i=1}^{2}\int_{0}^{\infty }dpp^{2}2\sqrt{%
E_{i}^{2}\left( p,M,b,l\right) }\text{ }
\end{equation}
with
\[
p^{2}+\frac{l\left( l+1\right) }{r_{0}^{2}}-\frac{3MG}{r_{0}^{3}}-\frac{3}{%
b^{2}}>0.
\]
For the SdS background we get
\begin{equation}
E\left( M,b\right) =\frac{V}{4\pi ^{2}}\sum_{l=0}^{\infty }\int_{0}^{\infty
}dpp^{2}\left( \sqrt{p^{2}+c_{-}^{2}}+\sqrt{p^{2}+c_{+}^{2}}\right)
\label{p37}
\end{equation}
where
\[
c_{\mp }^{2}=\frac{l\left( l+1\right) }{r_{0}^{2}}\mp \frac{3MG}{r_{0}^{3}}-%
\frac{3}{b^{2}},
\]
while when we refer to the dS space we put $M=0$ and $c^{2}=$ $l\left(
l+1\right) /r_{0}^{2}-3/b^{2}$. Here the meaning of the value $r_{0}$ is
that of maintaining the same boundary conditions to correctly compute the
Casimir-like energy. Then
\begin{equation}
E\left( b\right) =\frac{V}{4\pi ^{2}}\sum_{l=0}^{\infty }\int_{0}^{\infty
}dpp^{2}\left( 2\sqrt{p^{2}+c^{2}}\right)  \label{p38}
\end{equation}
Now, we are in position to compute the difference between $\left( \ref{p37}%
\right) $ and $\left( \ref{p38}\right) $. Since we are interested
in the ultraviolet limit, we have
\[
\Delta E\left( M,b\right) =E\left( M,b\right) -E\left( b\right)
\]
\[
=\frac{V}{4\pi ^{2}}\sum_{l=0}^{\infty }\int_{0}^{\infty }dpp^{2}\left[
\sqrt{p^{2}+c_{-}^{2}}+\sqrt{p^{2}+c_{+}^{2}}-2\sqrt{p^{2}+c^{2}}\right]
\]
\begin{equation}
=\frac{V}{4\pi ^{2}}\sum_{l=0}^{\infty }\int_{0}^{\infty }dpp^{3}\left[
\sqrt{1+\left( \frac{c_{-}}{p}\right) ^{2}}+\sqrt{1+\left( \frac{c_{+}}{p}%
\right) ^{2}}-2\sqrt{1+\left( \frac{c}{p}\right) ^{2}}\right]
\end{equation}
and for $p^{2}>>c_{\mp }^{2},c^{2}$, we obtain
\[
\frac{V}{4\pi ^{2}}\sum_{l=0}^{\infty }\int_{0}^{\infty }dpp^{3}\left[ 1+%
\frac{1}{2}\left( \frac{c_{-}}{p}\right) ^{2}-\frac{1}{8}\left( \frac{c_{-}}{%
p}\right) ^{4}+1+\frac{1}{2}\left( \frac{c_{+}}{p}\right) ^{2}-\frac{1}{8}%
\left( \frac{c_{+}}{p}\right) ^{4}\right.
\]
\begin{equation}
\left. -2-\left( \frac{c}{p}\right) ^{2}+\frac{1}{4}\left( \frac{c}{p}%
\right) ^{4}\right] =-\frac{V}{2\pi ^{2}}\frac{c_{M}^{4}}{8}\int_{0}^{\infty
}\frac{dp}{p},
\end{equation}
where $c_{M}^{2}=3MG/r_{0}^{3}$. We will use a cut-off $\Lambda $ to keep
under control the $UV$ divergence
\begin{equation}
\int_{0}^{\infty }\frac{dp}{p}\sim \int_{0}^{\frac{\Lambda }{c_{M}}}\frac{dx%
}{x}\sim \ln \left( \frac{\Lambda }{c_{M}}\right) ,
\end{equation}
where $\Lambda \leq m_{p}.$ Thus $\Delta E\left( M,b\right) $ for high
momenta becomes
\begin{equation}
\Delta E\left( M,b\right) \sim -\frac{V}{2\pi ^{2}}\frac{c_{M}^{4}}{16}\ln
\left( \frac{\Lambda ^{2}}{c_{M}^{2}}\right) =-\frac{V}{32\pi ^{2}}\left(
\frac{3MG}{r_{0}^{3}}\right) ^{2}\ln \left( \frac{r_{0}^{3}\Lambda ^{2}}{3MG}%
\right) .
\end{equation}
and Eq.$\left( \ref{i0}\right) $ to one loop is
\[
\left( E^{SdS}\left( M,b\right) -E^{dS}\left( b\right) \right) _{r\simeq
r_{0}>r_{+}}=-\frac{V}{32\pi ^{2}}\left( \frac{3MG}{r_{0}^{3}}\right)
^{2}\ln \left( \frac{r_{0}^{3}\Lambda ^{2}}{3MG}\right)
\]
\begin{equation}
=-\frac{V}{32\pi ^{2}}\left( \frac{3\left( r_{+}r_{++}\right) \left(
r_{+}+r_{++}\right) }{2\left( r_{+}^{2}+r_{+}r_{++}+r_{++}^{2}\right)
r_{0}^{3}}\right) ^{2}\ln \left( \frac{2\left(
r_{+}^{2}+r_{+}r_{++}+r_{++}^{2}\right) r_{0}^{3}\Lambda ^{2}}{3\left(
r_{+}r_{++}\right) \left( r_{+}+r_{++}\right) }\right) ,  \label{p38a}
\end{equation}
where we have used Eq.$\left( \ref{p12a}\right) $. On the other hand, when $%
r\longrightarrow r_{++}$%
\begin{equation}
x\left( r\right) \simeq \sqrt{2\kappa _{++}\left( r_{++}-r\right) }\qquad
V_{l}^{\mp }\left( x\right) \longrightarrow \frac{l\left( l+1\right) }{%
r_{++}^{2}}\mp \frac{3MG}{r_{++}^{3}}=const,
\end{equation}
where
\begin{equation}
\kappa _{++}=\lim\limits_{r\rightarrow r_{++}}\frac{1}{2}\left|
g_{00}^{\prime }\left( r\right) \right| =\frac{\left( r_{++}-r_{-}\right)
\left( r_{++}-r_{+}\right) }{2b^{2}r_{++}}  \label{p34a}
\end{equation}
is the ``outer'' surface gravity associated with the largest root. By
repeating the steps going from Eq.$\left( \ref{p36}\right) $ to Eq.$\left(
\ref{p38a}\right) $ with
\begin{equation}
E_{1,2}^{2}\left( p,M,b,l\right) =p^{2}+\frac{l\left( l+1\right) }{r_{++}^{2}%
}\mp \frac{3MG}{r_{++}^{3}}-\frac{3}{b^{2}},
\end{equation}
we obtain
\[
\left( E^{SdS}\left( M,b\right) -E^{dS}\left( b\right) \right) _{r\simeq
r_{++}}=-\frac{V}{32\pi ^{2}}\left( \frac{3MG}{r_{++}^{3}}\right) ^{2}\ln
\left( \frac{r_{++}^{3}\Lambda ^{2}}{3MG}\right)
\]
\begin{equation}
=-\frac{V}{32\pi ^{2}}\left( \frac{3r_{+}\left( r_{+}+r_{++}\right) }{%
2\left( r_{+}^{2}+r_{+}r_{++}+r_{++}^{2}\right) r_{++}^{2}}\right) ^{2}\ln
\left( \frac{2\left( r_{+}^{2}+r_{+}r_{++}+r_{++}^{2}\right)
r_{++}^{2}\Lambda ^{2}}{3r_{+}\left( r_{+}+r_{++}\right) }\right) .
\end{equation}
Like the Schwarzschild and the S-AdS cases, we observe that
\begin{equation}
\lim_{M\rightarrow 0}\lim_{r_{0}\rightarrow r_{+}}\Delta E\left( M,b\right)
\neq \lim_{r_{0}\rightarrow r_{+}}\lim_{M\rightarrow 0}\Delta E\left(
M,b\right) .
\end{equation}
This behavior seems to confirm that quantum effects come into play when we
try to reach the inner throat. Differently to the Schwarzschild and S-AdS
cases, here we have two scales corresponding to the two throats (horizons)
of the metric. This is an artifact of the approximation we have adopted to
deal with the differential equations of the system $\left( \ref{p34}\right) $%
. By defining a scale variable $x=3MG/\left( r_{0}^{3}\Lambda ^{2}\right) $
and a scale variable $y=3MG/\left( r_{++}^{3}\Lambda ^{2}\right) $, we
obtain
\[
E^{SdS}\left( M,b\right) -E^{dS}\left( b\right) =\left( E^{SdS}\left(
M,b\right) -E^{dS}\left( b\right) \right) _{r\simeq r_{0}>r_{+}}+\left(
E^{SdS}\left( M,b\right) -E^{dS}\left( b\right) \right) _{r\simeq r_{++}}
\]
\[
=-\frac{V}{32\pi ^{2}}\left[ \left( \frac{3MG}{r_{0}^{3}}\right) ^{2}\ln
\left( \frac{r_{0}^{3}\Lambda ^{2}}{3MG}\right) +\left( \frac{3MG}{r_{++}^{3}%
}\right) ^{2}\ln \left( \frac{r_{++}^{3}\Lambda ^{2}}{3MG}\right) \right]
\]
\begin{equation}
=\frac{V\Lambda ^{4}}{32\pi ^{2}}\left[ x^{2}\ln x+y^{2}\ln y\right] =\Delta
E\left( x,y\right) .
\end{equation}
A stationary point is reached for $x=y=0$, namely the dS space and another
stationary point is in $x=y=e^{-\frac{1}{2}}.$ This last one represents a
minimum of $\Delta E\left( x,y\right) $. This means that there is a
probability that the dS spacetime will be subjected to a topology change and
it will produce a SdS wormhole with a black hole pair\footnote{%
See also \cite{NoOd} for an approach of black hole pair creation
based on the effective action.} generated on the hypersurface
$\Sigma $. To see if this is really possible, we have to establish
if there exist unstable modes.

\section{Searching for negative modes}

\label{p4}

In this paragraph we look for negative modes of the eigenvalue equation $%
\left( \ref{p3a}\right) $. To this purpose we restrict the analysis to the S
wave. Indeed, in this state the centrifugal term is absent and this gives
the function $V\left( x\right) $ a potential well form, which is different
when the angular momentum $l\geq 1$. Moreover the potential well appears
only for the $H$ component, whose eigenvalue equation is
\begin{equation}
\left( -\triangle -\frac{4MG}{r^{3}}-\frac{2}{b^{2}}\right) H\left( r\right)
=-E_{0}^{2}H\left( r\right) .  \label{p40}
\end{equation}
$\triangle $ is the operator $\triangle _{l}$ of Eq.$\left( \ref{p33a}%
\right) $ with $l=0$ and $E_{0}^{2}>0$. By defining the reduced field $%
h\left( r\right) =H\left( r\right) r$, Eq.$\left( \ref{p40}\right) $ becomes
\begin{equation}
-\frac{d}{dr}\left( \sqrt{1-\frac{2MG}{r}-\frac{r^{2}}{b^{2}}}\frac{dh}{dr}%
\right) +\left( \frac{-3MG}{r^{3}}+\tilde{E}^{2}\right) \frac{h}{\sqrt{1-%
\frac{2MG}{r}-\frac{r^{2}}{b^{2}}}}=0,  \label{p41}
\end{equation}
where $\tilde{E}^{2}=-3/b^{2}+E_{0}^{2}$. By means of Eq.$\left( \ref{p13}%
\right) $, one gets
\[
-\frac{dx}{dr}\frac{d}{dx}\left( \sqrt{1-\frac{2MG}{r}-\frac{r^{2}}{b^{2}}}%
\frac{dh}{dx}\frac{dx}{dr}\right) +\left( -\frac{3MG}{r^{3}}+\tilde{E}%
^{2}\right) \frac{h}{\sqrt{1-\frac{2MG}{r}-\frac{r^{2}}{b^{2}}}}
\]
\begin{equation}
=-\frac{d}{dx}\left( \frac{dh}{dx}\right) +\left( -\frac{3MG}{r^{3}}+\tilde{E%
}^{2}\right) h=0.  \label{p42}
\end{equation}
Near the throat
\begin{equation}
x\left( r\right) \simeq \pm \frac{\sqrt{2r_{+}}}{\sqrt{\kappa _{+}}}\sqrt{%
\left( \frac{r}{r_{+}}-1\right) },
\end{equation}
where we have used Eq.$\left( \ref{p34b}\right) $ defining the inner surface
gravity. By defining the dimensionless variable $\rho =r/r_{+}$, we obtain $%
\rho \simeq 1+y^{2}$ where
\begin{equation}
y=\sqrt{\kappa _{+}}x/\sqrt{2r_{+}}=\tilde{\kappa}_{+}x.  \label{p43}
\end{equation}
Eq.$\left( \ref{p42}\right) $ becomes
\[
-\frac{d}{dy}\left( \frac{dh}{dy}\right) \tilde{\kappa}_{+}^{2}+\left( -%
\frac{3MG}{r_{+}^{3}\rho ^{3}\left( y\right) }+\tilde{E}^{2}\right) h
\]
\begin{equation}
=-\frac{d^{2}h}{dy^{2}}+\left( -\frac{3MG}{\tilde{\kappa}_{+}^{2}r_{+}^{3}%
\left( 1+y^{2}\right) ^{3}}+\lambda \right) h=0,
\end{equation}
where $\lambda =\tilde{E}^{2}/\tilde{\kappa}_{+}^{2}$. Expanding the
potential around $y=0$, one gets
\begin{equation}
-\frac{d^{2}h}{dy^{2}}+\left( -\frac{3MG}{\tilde{\kappa}_{+}^{2}r_{+}^{3}}%
\left( 1-3y^{2}\right) +\lambda \right) h
\end{equation}
\begin{equation}
=-\frac{d^{2}h}{dy^{2}}+\left( \omega ^{2}y^{2}-\frac{3MG}{\tilde{\kappa}%
_{+}^{2}r_{+}^{3}}+\lambda \right) h=0,
\end{equation}
with $\omega =\sqrt{9MG/\left( \tilde{\kappa}_{+}^{2}r_{+}^{3}\right) }$. In
this approximation we have obtained the equation of a quantum harmonic
oscillator equation whose spectrum is $E_{n}=\hbar \omega \left( n+\frac{1}{2%
}\right) $. Since we are using natural units, $\hbar $ is set to one and
\begin{equation}
\lambda _{n}=3MG/\left( \tilde{\kappa}_{+}^{2}r_{+}^{3}\right) -\sqrt{%
9MG/\left( \tilde{\kappa}_{+}^{2}r_{+}^{3}\right) }\left( n+\frac{1}{2}%
\right) .
\end{equation}
After some algebraic calculation, with the help of relation $\left( \ref
{p12a}\right) $, we obtain
\begin{equation}
\lambda _{n}=3\sqrt{\frac{2r_{++}\left( r_{+}+r_{++}\right) }{\left(
2r_{+}+r_{++}\right) \left( r_{++}-r_{+}\right) }}\left( \sqrt{\frac{%
2r_{++}\left( r_{+}+r_{++}\right) }{\left( 2r_{+}+r_{++}\right) \left(
r_{++}-r_{+}\right) }}-\frac{1}{\sqrt{2}}\left( n+\frac{1}{2}\right) \right)
.
\end{equation}
Let us examine the first two eigenvalues. The first one is
\begin{equation}
\lambda _{0}=3\sqrt{\frac{2r_{++}\left( r_{+}+r_{++}\right) }{\left(
2r_{+}+r_{++}\right) \left( r_{++}-r_{+}\right) }}\left( \sqrt{\frac{%
2r_{++}\left( r_{+}+r_{++}\right) }{\left( 2r_{+}+r_{++}\right) \left(
r_{++}-r_{+}\right) }}-\frac{1}{2}\right) .
\end{equation}
Since the eigenvalue must be positive, the following inequality must hold
\begin{equation}
\sqrt{\frac{2r_{++}\left( r_{+}+r_{++}\right) }{\left( 2r_{+}+r_{++}\right)
\left( r_{++}-r_{+}\right) }}>\frac{1}{2}\Longrightarrow \frac{7}{4}%
r_{+}r_{++}+\frac{7}{4}r_{++}^{2}+\frac{1}{2}r_{+}^{2}>0,
\end{equation}
which is verified $\forall \theta \in \left[ \frac{\pi }{2},\pi \right) $.
To proof that there is only one eigenvalue, we look at the second one
\begin{equation}
\lambda _{1}=3\sqrt{\frac{2r_{++}\left( r_{+}+r_{++}\right) }{\left(
2r_{+}+r_{++}\right) \left( r_{++}-r_{+}\right) }}\left( \sqrt{\frac{%
2r_{++}\left( r_{+}+r_{++}\right) }{\left( 2r_{+}+r_{++}\right) \left(
r_{++}-r_{+}\right) }}-\frac{3}{2}\right) .
\end{equation}
This implies the inequality
\begin{equation}
18r_{+}^{2}>r_{+}r_{++}+r_{++}^{2},
\end{equation}
which is verified when $\theta \in \left( 2.1708,\pi \right) $. Thus we have
only one eigenvalue when $\theta \in \left[ \frac{\pi }{2},2.1708\right) $.
In terms of $E^{2}$ we get
\begin{equation}
E^{2}=-3/b^{2}-\frac{3r_{++}\left( r_{+}+r_{++}\right) }{2b^{2}r_{+}^{2}}+%
\frac{3}{8b^{2}r_{+}^{2}}\sqrt{2r_{++}\left( r_{+}+r_{++}\right) \left(
2r_{+}+r_{++}\right) \left( r_{++}-r_{+}\right) }.
\end{equation}
Note that if we repeat the same calculation for the outer throat, i.e. $%
r_{++}$, we discover the absence of negative eigenvalues. Thus we can
conclude that there is only {\bf one eigenvalue} with the restriction that $%
\theta \in \left[ \frac{\pi }{2},2.1708\right) $. According to
Coleman \cite {Coleman}, this is a signal of a transition from a
false vacuum to a true one. It is evident that when we consider
the limit where $r_{++}\rightarrow r_{+}$, an infinite number of
eigenvalues enter in the discrete spectrum. This is a consequence
of the approximated parabolic potential which is used to describe
this extreme situation. To better deal with this problem, we
introduce the Nariai metric.

\section{The Nariai metric spin 2 operator and the evaluation of the energy
density}

\label{N}
\begin{figure}[tbh]
\vbox{\hfil\epsfxsize=4.5cm\epsfbox{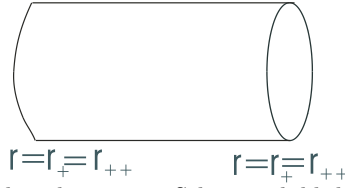}} \caption{The
geometry of the constant time slice degenerate Schwarzschild-de
Sitter (Nariai) spacetime with a polar angle suppressed. Isometric
copies of
this surface can be smoothly joined at the throats, producing a periodic $%
S^{2}\times S^{1}$ spatial topology.}
\label{f3}
\end{figure}

When $r_{+}=r_{++}$, the metric becomes degenerate and the function
$f\left( r\right) $ of Eq.$\left( \ref{p12}\right) $ becomes
\begin{equation}
f_{e}\left( r\right) =-\frac{1}{rb^{2}}\left( r-\bar{r}\right) ^{2}\left( r+2%
\bar{r}\right) ,
\end{equation}
where $\bar{r}=b/\sqrt{3}$. Since $f_{e}\left( r\right) \leq 0$ everywhere $r
$ becomes a time coordinate and $t$ becomes spatial. Nevertheless, this is
an artifact of a poor coordinate choice. To see what happens, we follow Ref.
\cite{GP} and we let $9M^{2}G^{2}\Lambda =1-3\varepsilon ^{2}$ so that the
limit $r_{+}\rightarrow r_{++}$ corresponds to $\varepsilon \rightarrow 0$.
We define a new radial coordinate $\theta _{1}$, and a new time coordinate $%
\phi _{1}$, by
\begin{equation}
\cos \theta _{1}=\frac{\sqrt{3}}{b\varepsilon }\left( r-\bar{r}\right)
\qquad \qquad \phi _{1}=t\frac{\sqrt{3}}{b\varepsilon },
\end{equation}
such that $r_{+}=\bar{r}-\varepsilon \sqrt{3}/b$ and $r_{++}=\bar{r}%
+\varepsilon \sqrt{3}/b$. To first order in $\varepsilon $ the metric
assumes the form
\[
ds^{2}=-\frac{b^{2}}{3}\left( 1+\frac{2}{3}\varepsilon \cos \theta
_{1}\right) \sin ^{2}\theta _{1}d\phi _{1}^{2}+\frac{b^{2}}{3}\left( 1-\frac{%
2}{3}\varepsilon \cos \theta _{1}\right) d\theta _{1}^{2}
\]
\begin{equation}
+\frac{b^{2}}{3}\left( 1-2\varepsilon \cos \theta _{1}\right) d\Omega ^{2},
\label{n1}
\end{equation}
describing a nearly degenerate SdS metric with two distinct roots. The
related surface gravities, to first order in $\varepsilon $ assume the
expressions \cite{BoHaw}
\begin{equation}
\kappa _{+,++}=\frac{b}{\sqrt{3}}\left( 1\mp \frac{2}{3}\varepsilon \cos
\theta _{1}\right) ,
\end{equation}
where the upper (lower) sign is for the cosmological (black hole) horizon.
When $\varepsilon \rightarrow 0$, the line element is
\begin{equation}
ds^{2}=\frac{b^{2}}{3}\left( -\sin ^{2}\theta _{1}d\phi _{1}^{2}+d\theta
_{1}^{2}+d\Omega ^{2}\right)
\end{equation}
and the surface gravities degenerate into one with the value
\begin{equation}
\kappa =\frac{b}{\sqrt{3}}.
\end{equation}
This is the Nariai metric\cite{Nariai} with topology $H^{2}\times S^{2}$.
Its Euclidean form has the well known topology $S^{2}\times S^{2}$ and form
\begin{equation}
ds^{2}=\frac{b^{2}}{3}\left( d\Omega _{1}^{2}+d\Omega ^{2}\right) .
\end{equation}
Every constant time section has the topology $S^{1}\times S^{2}$ with the
throats of the same size as illustrated in Fig.\ref{f3}

However we are interested for the Lorentzian version. In this case,
Regge-Wheeler decomposition shows that the even-parity three-dimensional
perturbation is
\begin{equation}
h_{ij}^{even}\left( \theta _{1},\theta ,\phi \right) =\frac{b^{2}}{3}diag%
\left[ H\left( \theta _{1}\right) ,K\left( \theta _{1}\right) ,\sin
^{2}\theta K\left( \theta _{1}\right) \right] Y_{l0}\left( \theta ,\phi
\right) .
\end{equation}
For a generic value of the angular momentum $L$, one gets
\begin{equation}
\left\{
\begin{array}{c}
\frac{3}{b^{2}}\left( -\partial _{\theta _{1}}^{2}+l\left( l+1\right)
-2\right) H\left( \theta _{1}\right) =E_{l}^{2}H\left( \theta _{1}\right) \\
\\
\frac{3}{b^{2}}\left( -\partial _{\theta _{1}}^{2}+l\left( l+1\right)
\right) K\left( \theta _{1}\right) =E_{l}^{2}K\left( \theta _{1}\right)
\end{array}
\right. ,
\end{equation}
where the associated three dimensional mixed Ricci tensor is
\begin{equation}
R_{i}^{j}=\left\{ 0,\frac{3}{b^{2}},\frac{3}{b^{2}}\right\} .
\end{equation}
Also in this case the unstable mode appears for the $l=0$ case leading to
the eigenvalue equation
\begin{equation}
\left( -\partial _{\theta _{1}}^{2}-2\right) H\left( \theta _{1}\right)
=-\lambda H\left( \theta _{1}\right) ,
\end{equation}
where $\lambda =b^{2}E^{2}/3.$ The eigenvalue is easily determined and its
value is
\begin{equation}
\lambda =2,
\end{equation}
with eigenfunction
\begin{equation}
H\left( \theta _{1}\right) =const.
\end{equation}
Since the range of integration is finite due to the periodicity of the
argument, the eigenfunction is normalizable. For the other component we have
no solutions at all, because the operator is the same of a free particle
having only a continuous spectrum. For completeness, we calculate the energy
difference between the SdS background in the Nariai form and the dS metric
in the stable sector. The total energy in the presence of the Nariai metric
is
\begin{equation}
E_{Nariai}\left( b\right) =\frac{V}{2\pi ^{2}}\frac{1}{2}\sum_{l=0}^{\infty
}\int_{0}^{\infty }dpp^{2}\left( \sqrt{p^{2}+\frac{3l\left( l+1\right) }{%
b^{2}}-6/b^{2}}+\sqrt{p^{2}+\frac{3l\left( l+1\right) }{b^{2}}}\right) ,
\end{equation}
while for the pure de Sitter metric, we have
\begin{equation}
E_{dS}\left( b\right) =\frac{V}{2\pi ^{2}}\frac{1}{2}\sum_{l=0}^{\infty
}\int_{0}^{\infty }dpp^{2}2\left( \sqrt{p^{2}+\frac{3l\left( l+1\right) }{%
b^{2}}-3/b^{2}}\right) .
\end{equation}
The Casimir-like energy becomes
\[
\Delta E\left( b\right) =E_{Nariai}\left( b\right) -E_{dS}\left( b\right)
\]
\[
=\frac{V}{2\pi ^{2}}\frac{1}{2}\sum_{l=0}^{\infty }\int_{0}^{\infty }dpp^{2}%
\left[ \sqrt{p^{2}+\frac{3l\left( l+1\right) }{b^{2}}-6/b^{2}}+\sqrt{p^{2}+%
\frac{3l\left( l+1\right) }{b^{2}}}\right.
\]
\begin{equation}
\left. -2\sqrt{p^{2}+\frac{3l\left( l+1\right) }{b^{2}}-3/b^{2}}\right]
\end{equation}
When $p^{2}>>6/b^{2}$ we write
\begin{equation}
\Delta E\left( b\right) \simeq -\frac{V}{2\pi ^{2}}\frac{9}{4b^{4}}%
\int_{0}^{\infty }\frac{dp}{p}.
\end{equation}
Introducing an $UV$ cut-off one gets
\begin{equation}
\int_{0}^{\infty }\frac{dp}{p}\sim \int_{0}^{\Lambda ^{2}b^{2}}\frac{dx}{x}%
\sim \ln \left( \Lambda ^{2}b^{2}\right)
\end{equation}
and $\Delta E\left( b\right) $ for high momenta becomes
\begin{equation}
\Delta E\left( b\right) \sim -\frac{V}{2\pi ^{2}}\frac{9}{16b^{4}}\ln \left(
\Lambda ^{2}b^{2}\right) =-\frac{V}{32\pi ^{2}}\frac{9}{b^{4}}\ln \left(
\Lambda ^{2}b^{2}\right) .
\end{equation}

\section{Boundary Reduction and stability}

\label{p5}An equivalent approach to Eq.$\left( \ref{p41}\right) $ can be set
up by means of a variational procedure applied on a functional whose minimum
represents the solution of the problem. Let us define
\[
J\left( h,E^{2}\right) =\frac{1}{2}\int\limits_{0}^{\bar{x}}dx\left[ \left(
\frac{dh\left( x\right) }{dx}\right) ^{2}-\frac{3MG}{r^{3}\left( x\right) }%
h^{2}\left( x\right) \right] +\frac{\tilde{E}^{2}}{2}\int\limits_{0}^{\bar{x}%
}h^{2}\left( x\right) dx,
\]
where $dx$ is given by Eq.$\left( \ref{p11a}\right) $. Eq.$\left( \ref{p41}%
\right) $ is equivalent to find the minimum of
\begin{equation}
\tilde{E}^{2}=\frac{\int\limits_{0}^{\bar{x}}dx\left[ \left( \frac{dh\left(
x\right) }{dx}\right) ^{2}-\frac{3MG}{r^{3}\left( x\right) }h^{2}\left(
x\right) \right] }{\int\limits_{0}^{\bar{x}}h^{2}\left( x\right) dx}.
\label{p51}
\end{equation}
For future purposes, we use the boundary conditions
\begin{equation}
h\left( \bar{x}\right) =0.
\end{equation}
When $2MG/r\ll 1$ Eq.$\left( \ref{p42}\right) $ becomes
\begin{equation}
-\frac{d^{2}h}{dx^{2}}+\tilde{E}^{2}h=0  \label{p51a}
\end{equation}
and the solution is
\begin{equation}
h\left( x\right) =A\exp \left( -\tilde{E}x\right) +B\exp \left( \tilde{E}%
x\right) .
\end{equation}
In the approximation of Eq.$\left( \ref{p51a}\right) $, we can suppose $x$
so large that the increasing exponential has to be eliminated in such a way
to take only $h\left( x\right) =A\exp \left( -\tilde{E}x\right) $. If we
change the variables in a dimensionless form like Eq.$\left( \ref{p43}%
\right) $, we get
\begin{equation}
\mu =\frac{\tilde{E}^{2}}{\tilde{\kappa}^{2}}=\frac{\int\limits_{0}^{\bar{y}%
}dy\left[ \left( \frac{dh\left( y\right) }{dy}\right) ^{2}-\frac{3MG}{%
r_{+}^{3}\tilde{\kappa}_{+}^{2}\rho ^{3}\left( y\right) }h^{2}\left(
y\right) \right] }{\int\limits_{0}^{y\left( a\right) }dyh^{2}\left( y\right)
}.  \label{p52}
\end{equation}
The asymptotic behaviour of $h\left( x\right) $ suggests to choose $h\left(
\lambda ,y\right) =\exp \left( -\lambda y\right) $ as a trial function, and
Eq.$\left( \ref{p52}\right) $ becomes
\begin{equation}
\mu \left( \lambda \right) =\lambda ^{2}-\frac{3MG}{r_{+}^{3}\tilde{\kappa}%
_{+}^{2}}\frac{\int\limits_{0}^{\bar{y}}\frac{dy}{\rho ^{3}\left( y\right) }%
\exp \left( -2\lambda y\right) }{\frac{1-\exp \left( -2\lambda y\right) }{%
2\lambda }}.
\end{equation}
Close to the throat $\exp \left( -2\lambda y\right) \simeq 1-2\lambda y$ and
\begin{equation}
\mu \left( \lambda \right) =\lambda ^{2}-\frac{3MG}{r_{+}^{3}\tilde{\kappa}%
_{+}^{2}}+\frac{9MG}{r_{+}^{3}\tilde{\kappa}_{+}^{2}}\left[ \frac{\bar{y}}{%
2\lambda }+\bar{y}^{2}\right] .
\end{equation}
The minimum of $\mu \left( \lambda \right) $ is reached for $\bar{\lambda}%
=\left( \frac{9MG}{4r_{+}^{3}\tilde{\kappa}_{+}^{2}}\bar{y}\right) ^{\frac{1%
}{3}}$ assuming therefore the value
\begin{equation}
\mu \left( \bar{\lambda}\right) =3\left( \frac{9}{4}D\bar{y}\right) ^{\frac{2%
}{3}}-3D+3D\bar{y}^{2},  \label{p53}
\end{equation}
where
\[
D=\frac{MG}{r_{+}^{3}\tilde{\kappa}_{+}^{2}}=\frac{2r_{++}\left(
r_{+}+r_{++}\right) }{\left( 2r_{+}+r_{++}\right) \left( r_{++}-r_{+}\right)
}
\]
\begin{equation}
=\frac{2\cos \left( \frac{\theta }{3}\right) \left( \cos \left( \frac{\theta
+4\pi }{3}\right) +\cos \left( \frac{\theta }{3}\right) \right) }{\left(
2\cos \left( \frac{\theta +4\pi }{3}\right) +\cos \left( \frac{\theta }{3}%
\right) \right) \left( \cos \left( \frac{\theta }{3}\right) -\cos \left(
\frac{\theta +4\pi }{3}\right) \right) }.  \label{p54}
\end{equation}
If we consider the value of $\theta $ such that only one eigenvalue appears,
we obtain $D=2.25$ and Eq.$\left( \ref{p53}\right) $becomes zero for $\bar{y}%
_{c}=.463\,68$ corresponding to $\bar{\rho}_{c}=1.215$. This means that the
unstable mode persists until the boundary radius $\bar{\rho}$ falls below $%
\bar{\rho}_{c}$.

\section{Summary and Conclusions}

\label{p6}In this paper we have extended the computation of the Casimir-like
energy to the case of the Schwarzschild-de Sitter (SdS) background with the
de Sitter (dS) space as a reference space. This evaluation has been done to
one-loop in the TT (transverse, traceless) sector which is the gauge
invariant part of the quantum fluctuation of the gravitational field. As
stressed in the introduction to correctly compute the Casimir energy we need
to subtract field configurations which have the same asymptotic properties
and the same asymptotic boundary conditions. For the SdS metric and the dS
metric, this is the case. In this context a lot of work has been done; for
example in Refs.\cite{GP,Young,VW}, it has been shown the existence of one
negative mode in the TT sector when the saddle point approximation is
considered: a clear sign of instability. In particular, the instability has
been related to the probability of creating a black hole pair\cite{VW,Remo3}%
. However this particular result has been obtained by looking at the
partition function and therefore with the introduction of an equilibrium
temperature, that in the case of the pure Schwarzschild and flat metric can
be imposed to be equal, but in the present case (i.e. the SdS metric and the
dS metric) it cannot. Therefore as stressed in Ref.\cite{VW}, it is not very
clear how these spaces having a different temperature (periodicity) can be
compared. On the other hand if we adopt the hamiltonian approach we can
avoid the introduction of a temperature and it is possible to build a scheme
where the classical contribution is conserved; this point is fundamental to
discuss the instability\cite{Witten}. What we have found, in our framework,
is the well known existence of an unstable mode in the S wave for the
extreme SdS metric (Nariai)\footnote{%
For this last point a discussion can be found in
Ref.\cite{Remo3}.} but we have also found that an unstable mode
exists also for the SdS metric. To interpret such an instability
as a decay process nucleating a black hole pair, we need to show
the existence of only ``{\bf one negative mode}'' \cite {Coleman}.
Unfortunately, in our approximation we have discovered a
dependence of the number of negative modes on $\theta $ the
variable defined in Eq.$\left( \ref{p11}\right) $. In particular,
we have seen that the closest is the approach to the extreme value
$\theta =\pi $, the highest is the number of negative eigenvalues
falling into the negative spectrum. This abundance of negative
eigenvalues can be limited if we restrict the
variability of $\theta $ into the interval $\left[ \frac{\pi }{2}%
,2.1708\right) $; in this range Coleman arguments can be applied.
However the proliferation of negative modes in the nearly extremal
SdS metric is a consequence of the approximated potential of the
eigenvalue equation. Indeed by using nearly degenerate coordinates
like those in Eq.$\left( \ref{n1}\right) $, only one negative
eigenvalue appears. It is interesting to observe that the
appearance of a negative mode, even for the SdS metric, can be
related with the production of a sub-maximal black hole
pair\cite{BoHaw1}. What is interesting to observe is the existence
of a critical radius $\rho _{c}$ below which the instability
disappears. This could open the possibility of an existing
foam-like space composed by copies of bubbles in analogy with the
model discussed in Ref.\cite{Remo4}. Even in this case, a
dependence of the ultraviolet cut-off is present. This is
principally due to the non renormalizability of quantum gravity.
However, the fact that the same divergent behaviour appears also
in this case it is a signal of a more general situation typically
concerning the spherically symmetric metrics, that is every
spherically symmetric metric describing a wormhole (black hole)
can produce in the spectrum of quantum fluctuations in a
semiclassical approximation one negative mode, provided the
boundary conditions be energy conserving.

\section{Acknowledgements}

The author would like to thank R. Bousso and M. Volkov for useful
comments and suggestions.

\appendix

\section{Kruskal-Szekeres coordinates for SdS spacetime}

\label{app1}

\begin{figure}[tbh]
\vbox{\hfil\epsfxsize=6.5cm\epsfbox{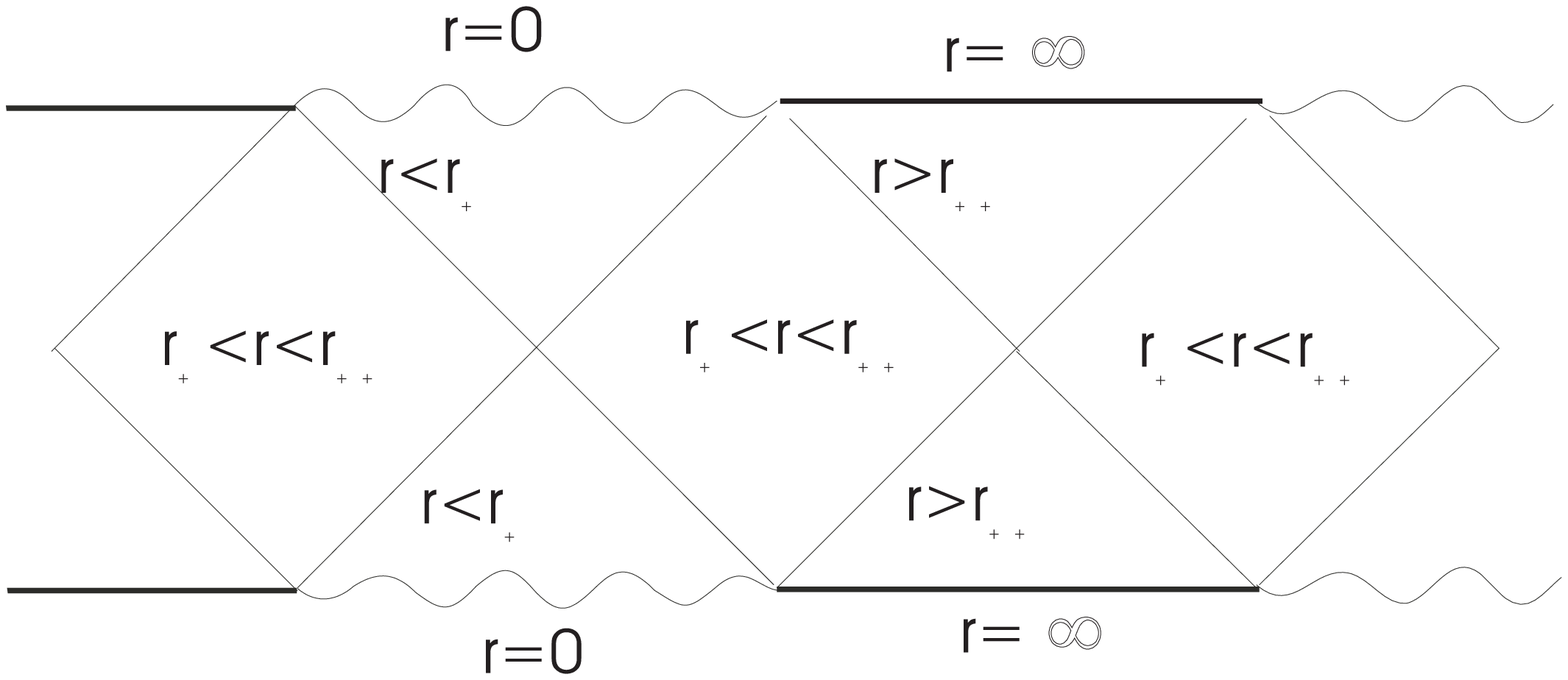}\hfil}
\caption{Penrose diagram for the Schwarzschild-de Sitter
spacetime.} \label{f4}
\end{figure}

We have defined the SdS line element in Eq.$\left( \ref{p11aa}\right) $. To
introduce the Kruskal-Szekeres\cite{KS,HawEll,MTW} type coordinates we
consider the following transformation
\[
ds^{2}=-\left( 1-\frac{2MG}{r}-\frac{r^{2}}{b^{2}}\right) \left[
dt^{2}-dr^{\ast 2}\right] +r^{2}d\Omega ^{2}
\]
\begin{equation}
=-\left( 1-\frac{2MG}{r}-\frac{r^{2}}{b^{2}}\right) dvdu+r^{2}\left(
u,v\right) d\Omega ^{2},  \label{a1}
\end{equation}
where $v=t+r^{\ast }$ is the ingoing radial null coordinate and $u=t-r^{\ast
}$ is the outgoing radial null coordinate. The ``tortoise coordinate'' $%
r^{\ast }$ is defined by
\begin{equation}
dr^{\ast }=-\frac{rb^{2}dr}{\left( r-r_{+}\right) \left( r-r_{++}\right)
\left( r+r_{+}+r_{++}\right) }
\end{equation}
and by means of the surface gravity associated to each root, we write
\begin{equation}
\frac{1}{f}=\frac{1}{2\kappa _{+}\left( r-r_{+}\right) }+\frac{1}{2\kappa
_{++}\left( r-r_{++}\right) }+\frac{1}{2\kappa _{-}\left(
r+r_{+}+r_{++}\right) },
\end{equation}
where $f\left( r\right) $ has been defined by Eq.$\left( \ref{p11a}\right) $%
, while $\kappa _{++}$ and $\kappa _{+}$ by Eq.$\left( \ref{p34a}\right) $
and Eq.$\left( \ref{p34b}\right) $, respectively. The last surface gravity
associated with the negative root is
\begin{equation}
\kappa _{-}=\frac{\left( 2r_{+}+r_{++}\right) \left( 2r_{++}+r_{+}\right) }{%
2b^{2}\left( r_{+}+r_{++}\right) }.
\end{equation}
Thus
\begin{equation}
r^{\ast }=\frac{1}{2\kappa _{+}}\ln \left| \frac{r}{r_{+}}-1\right| +\frac{1%
}{2\kappa _{++}}\ln \left| \frac{r}{r_{++}}-1\right| +\frac{1}{2\kappa _{-}}%
\ln \left( \frac{r}{r_{+}+r_{++}}+1\right)
\end{equation}
To avoid singularities we can define Kruskal-Szekeres type coordinates
\begin{equation}
V^{++}=\exp \kappa _{++}v\qquad U^{++}=-\exp -\kappa _{++}u.
\end{equation}
These coordinates do not cover $r\leq r_{+}$ because of the coordinate
singularity at $r=r_{+}$ (and $U^{++}V^{++}$ is complex for $r\leq r_{+}$),
but $r=r_{+}$ and a similar four regions are covered by the $\left(
U^{+},V^{+}\right) $ Kruskal-Szekeres-type coordinates to this case.
\begin{equation}
V^{+}=\exp \kappa _{+}v\qquad U^{+}=-\exp -\kappa _{+}u.
\end{equation}
For the $++$ sign we have
\begin{equation}
U^{++}V^{++}=-\exp \left( \kappa _{++}\left( v-u\right) \right) =-\exp
\left( 2\kappa _{++}r^{\ast }\right) =-\left( \frac{r}{r_{+}}-1\right) ^{%
\frac{\kappa _{++}}{\kappa _{+}}}\left( \frac{r}{r_{+}+r_{++}}+1\right) ^{%
\frac{\kappa _{++}}{\kappa -}}\left( \frac{r}{r_{++}}-1\right)
\end{equation}
and the respective line element is
\[
ds_{++}^{2}=-\frac{\left( r_{+}+r_{++}\right) r_{+}r_{++}}{b^{2}\kappa
_{++}^{2}r}\left( \frac{r}{r_{+}+r_{++}}+1\right) ^{1-\frac{\kappa _{++}}{%
\kappa -}}\left( \frac{r}{r_{+}}-1\right) ^{1-\frac{\kappa _{++}}{\kappa _{+}%
}}dU^{++}dV^{++}+r^{2}\left( U^{++},V^{++}\right) d\Omega ^{2}
\]
\begin{equation}
=-\frac{2MG}{\kappa _{++}^{2}r}\left( \frac{r}{r_{+}+r_{++}}+1\right) ^{1-%
\frac{\kappa _{++}}{\kappa -}}\left( \frac{r}{r_{+}}-1\right) ^{1-\frac{%
\kappa _{++}}{\kappa _{+}}}dU^{++}dV^{++}+r^{2}\left( U^{++},V^{++}\right)
d\Omega ^{2},
\end{equation}
while for the $+$ sign we have
\begin{equation}
U^{+}V^{+}=-\exp \left( \kappa _{+}\left( v-u\right) \right) =-\exp \left(
2\kappa _{+}r^{\ast }\right) =-\left( r-r_{+}\right) \left(
r+r_{+}+r_{++}\right) ^{\frac{\kappa _{+}}{\kappa -}}\left( r_{++}-r\right)
^{\frac{\kappa _{+}}{\kappa _{++}}}
\end{equation}
and the associated line element is
\[
ds_{+}^{2}=-\frac{\left( r_{+}+r_{++}\right) r_{+}r_{++}}{b^{2}\kappa
_{+}^{2}r}\left( \frac{r}{r_{+}+r_{++}}+1\right) ^{1-\frac{\kappa _{+}}{%
\kappa -}}\left( 1-\frac{r}{r_{++}}\right) ^{1-\frac{\kappa _{+}}{\kappa
_{++}}}dU^{+}dV^{+}+r^{2}\left( U^{+},V^{+}\right) d\Omega ^{2}
\]
\begin{equation}
=-\frac{2MG}{\kappa _{+}^{2}r}\left( \frac{r}{r_{+}+r_{++}}+1\right) ^{1-%
\frac{\kappa _{+}}{\kappa -}}\left( 1-\frac{r}{r_{++}}\right) ^{1-\frac{%
\kappa _{+}}{\kappa _{++}}}dU^{+}dV^{+}+r^{2}\left( U^{+},V^{+}\right)
d\Omega ^{2}.
\end{equation}
The conformal Penrose diagram of the SdS space is shown in Fig.\ref{f4}.
Regions limited by $r_{+}<r<r_{++}$ lie between the black hole and
cosmological horizon. Regions $r>r_{++}$ correspond to an asymptotic de
Sitter region and region $r_{+}>r$ to the black hole interior.

\end{document}